%% file: paper.tex
\documentclass{PoS}

\PoS{PoS(LAT2005)002}

\title{Lattice QCD with light Wilson quarks%
\thanks{Based in part on work done in collaboration with L.~Del Debbio,
L.~Giusti, R.~Petronzio and N.~Tantalo}}

\ShortTitle{Lattice QCD with light Wilson quarks}

\author{{Martin L\"uscher}\\

        CERN, Physics Department, TH Division, CH-1211 Geneva 23, Switzerland\\	

        E-mail: \email{luscher@mail.cern.ch}}

\usepackage{epsfig}
\usepackage{latexsym}

\abstract{
Wilson's formulation of lattice QCD is attractive for many reasons, but 
perhaps mainly because of its simplicity and conceptual clarity.
Numerical simulations of the Wilson theory (and of its improved versions) 
tend to be extremely demanding, however, to the extent that they rapidly 
become impractical at small quark masses.
Recent advances in simulation algorithms now allow 
such simulations to be pushed to significantly smaller
masses without having to compromise in other ways.
Contact with chiral perturbation theory can thus be made
and many physics questions can be addressed that previously
appeared to be inaccessible.
}

\FullConference{XXIIIrd International Symposium on Lattice Field Theory\\

		 25--30 July 2005\\

		 Trinity College, Dublin, Ireland}

\begin{document}

\input macros

\input sect1

\input sect2

\input sect3

\input sect4

\input sect5

\input sect6

\input sect7

\input sect8

\input sect9

\input sect10

\input sect11

\input biblio
\end{document}

%% file: macros


\def\rmd{{\rm d}}
\def\rmD{{\rm D}}
\def\rme{{\rm e}}
\def\rmO{{\rm O}}
\def\rmU{{\rm U}}
\def\rmo{{\rm o}}


\def\rz{{\Bbb R}}
\def\gz{{\Bbb Z}}
\def\nz{{\Bbb N}}
\def\Im{{\rm Im}\,}
\def\Re{{\rm Re}\,}


\def\defeq{\mathrel{\mathop=^{\rm def}}}
\def\proof{\noindent{\sl Proof:}\kern0.6em}
\def\endproof{\hskip0.6em plus0.1em minus0.1em
\setbox0=\null\ht0=5.4pt\dp0=1pt\wd0=5.3pt
\vbox{\hrule height0.8pt
\hbox{\vrule width0.8pt\box0\vrule width0.8pt}
\hrule height0.8pt}}
\def\frac#1#2{\hbox{$#1\over#2$}}
\def\dual{\mathstrut^*\kern-0.1em}
\def\mod{\;\hbox{\rm mod}\;}
\def\ring{\mathaccent"7017}
\def\lvec#1{\setbox0=\hbox{$#1$}
    \setbox1=\hbox{$\scriptstyle\leftarrow$}
    #1\kern-\wd0\smash{
    \raise\ht0\hbox{$\raise1pt\hbox{$\scriptstyle\leftarrow$}$}}
    \kern-\wd1\kern\wd0}
\def\rvec#1{\setbox0=\hbox{$#1$}
    \setbox1=\hbox{$\scriptstyle\rightarrow$}
    #1\kern-\wd0\smash{
    \raise\ht0\hbox{$\raise1pt\hbox{$\scriptstyle\rightarrow$}$}}
    \kern-\wd1\kern\wd0}
\def\slash#1{\setbox0=\hbox{$#1$}\setbox1=\hbox{$\kern1pt/$}
    #1\kern-\wd0\kern1pt/\kern-\wd1\kern\wd0}


\def\nab#1{{\nabla_{#1}}}
\def\nabstar#1{{\nabla\kern0.5pt\smash{\raise 4.5pt\hbox{$\ast$}}
               \kern-5.5pt_{#1}}}
\def\drv#1{{\partial_{#1}}}
\def\drvstar#1{{\partial\kern0.5pt\smash{\raise 4.5pt\hbox{$\ast$}}
               \kern-6.0pt_{#1}}}
\def\ldrv#1{{\lvec{\,\partial}_{#1}}}
\def\ldrvstar#1{{\lvec{\,\partial}\kern-0.5pt\smash{\raise 4.5pt\hbox{$\ast$}}
               \kern-5.0pt_{#1}}}


\def\MeV{{\rm MeV}}
\def\GeV{{\rm GeV}}
\def\TeV{{\rm TeV}}
\def\fm{{\rm fm}}


\def\euler{\gamma_{\rm E}}


\def\psibar{\overline{\psi}}
\def\chibar{\overline{\chi}}
\def\psitilde{\widetilde{\psi}}
\def\chileft{\chi_{\rm L}}
\def\chiright{\chi_{\rm R}}


\def\dirac#1{\gamma_{#1}}
\def\diracstar#1#2{
    \setbox0=\hbox{$\gamma$}\setbox1=\hbox{$\gamma_{#1}$}
    \gamma_{#1}\kern-\wd1\kern\wd0
    \smash{\raise4.5pt\hbox{$\scriptstyle#2$}}}
\def\dirachat{\hat{\gamma}_5}


\def\group{G}
\def\algebra{{\frak g}}
\def\SUtwo{{\rm SU(2)}}
\def\SUthree{{\rm SU(3)}}
\def\SUn{{\rm SU}(N)}
\def\tr{{\rm tr}}
\def\Tr{{\rm Tr}}
\def\Ad{{\rm Ad}\kern0.1em}
\def\Exp{{\rm E}}


\def\Dw{D_{\rm w}}
\def\Dhat{\hat{D}}
\def\Dslash{D\kern-1.5ex/\kern0.6ex}
\def\Msap{M_{\rm sap}}


\def\B{\Lambda}
\def\Bs{\B^{\kern-0.1em*}}
\def\dB{\partial\B}
\def\dBs{\partial\Bs}
\def\DB{D_{\B}}
\def\DBs{D_{\Bs_{\vphantom{1}}}}
\def\DdB{D_{\dB}}
\def\DdBs{D_{\dBs_{\vphantom{1}}}}

\def\Om{\Omega}
\def\Oms{\Omega^{\ast}}
\def\dOm{\partial\Om}
\def\dOms{\partial\Oms}
\def\DOm{D_{\Om}}
\def\DOms{D_{\Oms_{\vphantom{1}}}}
\def\DdOm{D_{\dOm}}
\def\DdOms{D_{\dOms_{\vphantom{1}}}}

\def\thetaB{\theta_{\B}}
\def\PrdB{\theta_{\dB}}


\def\ncy{n_{\rm cy}}
\def\nmr{n_{\rm mr}}
\def\nkv{n_{\rm kv}}


\def\mq{m}
\def\ms{m_{\rm s}}
\def\mpi{m_{\pi}}
\def\mrho{m_{\rho}}
\def\Wrho{W_{\rho}}
\def\Fpi{F_{\pi}}


\def\Nf{N_{\rm f}}


\def\SF{S_{\rm F}}
\def\SG{S_{\rm gauge}}
\def\Spf{S_{\rm pf}}
\def\Stot{S_{\rm tot}}
\def\Ftot{F}
\def\Fgauge{F_{\rm G}}
\def\Fblock{F_{\Lambda}}
\def\Fglobal{F_{R}}


\def\epsgauge{\epsilon_{0}}
\def\epsblock{\epsilon_{1}}
\def\epsglobal{\epsilon_{2}}

\def\vsp{{\vphantom{$a_b$}}}
\def\thicktablerule{\hrule height1pt}
\def\thintablerule{\hrule height0.4pt}

\renewcommand{\thefootnote}{\fnsymbol{footnote}}

%% file: sect1
\section{Introduction}

Several formulations of lattice QCD are currently in use, 
all having various advantages and shortcomings.
There are good reasons, however, to proceed with 
the one proposed by Wilson long ago \cite{Wilson}.
In particular, this formulation respects basic principles
such as locality and unitarity,
and it also preserves most symmetries of the continuum theory exactly.
The axial chiral symmetries are among those that are not preserved,
but it was understood in the 90's that the chiral-symmetry violations
can be reduced to small corrections of order $a^2$ 
(at lattice spacings $a\leq0.1\,\fm$) by adding O($a$) counterterms to
the lattice action and the composite gauge-invariant 
fields of interest \cite{SW,OaImp}.
Non-perturbative renormalization techniques were then developed, and 
many physical quantities were calculated  
in the quenched (or valence)
approximation, using numerical simulations.

When the sea-quark effects are included in the simulations,
much more computer time is required, particularly 
so in the chiral regime of QCD,
where the widely used simulation algorithms slow down
proportionally to the second or maybe even the third power of the 
masses of the light quarks
\cite{SesamTXLauto}--\cite{CPPACSsmall}.
So far, simulations of the Wilson theory 
(with or without O($a$) corrections)
thus proved to be impractical at 
lattice spacings $a\leq0.1\,\fm$ and light-quark masses significantly smaller
than half the physical mass of the strange quark.

%% file: sect2
\section{Recent advances in simulation algorithms}

Eventually many different lattices will have to be simulated,
with lattice spacings as small as $0.04$ fm perhaps, and
spatial sizes from, say, $2$ to $5$ fm.
Moreover, the light-quark masses on these lattices
should be such that the physical
point can be reached with at most a short extrapolation.
Taken together, these requirements imply large lattices,
of size $48\times24^3, 64\times32^3$ and $96\times48^3$ for example,
and a situation, where the numerical solution of the 
lattice Dirac equation for a single source field requires as much as
$10^3$--$10^4$ applications of the Dirac operator.

While the available computers become more powerful
at an exponential rate, it is quite clear that
better algorithms will be needed as well
if these simulations are to be performed in the next few years.
There are in fact
new opportunities for algorithmic improvements 
once very large lattices are considered.
The separation of short- and long-distance effects
along the lines proposed by Peardon and Sexton \cite{PeardonSexton}, 
for example, is potentially more profitable if the lattice
covers a wide range of scales.

The last few years have seen several developments 
that go in this general direction, 
one of them being the introduction of 
domain-decomposition methods in lattice QCD \cite{SchwarzAlgorithmI}.
This led to a fast parallel solver for the 
lattice Dirac equation \cite{SchwarzSolver} and
to an efficient
simulation algorithm for two-flavour QCD (with
mass-degenerate quarks), 
which scales well
with the lattice size and the quark mass
\cite{SchwarzAlgorithmII}.

Progress has also been made using the
Hybrid-Monte-Carlo (HMC) algorithm \cite{HMC}
with even--odd preconditioning
and two pseudo-fermion fields, as suggested some time ago by Hasenbusch
\cite{Hasenbusch}. Early studies of this algorithm
reported speed-up factors of $2$ or so 
with respect to the case where
only one pseudo-fermion field is used
\cite{HasenbuschJansen,DellaMorteEtAl},
but according to a recent paper by
Urbach et al.~\cite{UrbachEtAl}, more impressive 
acceleration factors can be achieved at small quark masses 
if the para\-meters of the algorithm are properly tuned.

In this talk, however, the focus will be on  
the new simulation algorithm mentioned above,
which is based on a domain decomposition of the lattice.
The general strategies are first discussed,
then the algorithm itself, and finally some
simulation results \cite{CHQCD}
that have been obtained with it.

%% file: sect3
\section{Domain-decomposition methods}

Probably the first domain-decomposition method goes back to Hermann
Amandus Schwarz, a well-known mathematician of the 19th century.
One of the questions Schwarz posed himself was whether it is possible
to solve the Dirichlet problem in complicated domains
(the union of a disk and a square for example) in a 
constructive way. He then showed that the solution can 
be obtained through a convergent iteration, in which
the Laplace equation is solved alternately in overlapping sub-domains,
with boundary 
values given by the current approximate solution in the full domain.
This ingenious method is now
referred to as the \emph{Schwarz alternating procedure} and is 
widely used, in the form of a preconditioner, for the numerical solution 
of elliptic partial differential equations 
(see ref.~\cite{Saad} for an introduction and a list of references).

There are at least three reasons 
for the popularity of domain-decomposition methods:

\vspace{0.2cm}\noindent
\emph{Parallel efficiency.} 
Domain decompositions often map to the nodes of a parallel computer
in a natural way. Using specially designed algorithms
that spend most of the time on the sub-domains, it is then possible
to reduce the communication overhead significantly. 
While this is a purely tech\-ni\-cal point, the parallel
efficiency of an algorithm is an all-important issue in
practice, when large systems are considered.

\vspace{0.2cm}\noindent
\emph{Mode separation.}
In many cases the dynamics of the low- and high-frequency modes of 
the system is very different so that it may be advisable to treat
these modes separately.
The division of space into sub-domains divides the modes 
into local and global ones, thus
providing an obvious opportunity to implement the mode separation.

\vspace{0.2cm}\noindent
\emph{Scaling behaviour.}
Good algorithms should scale slowly with the parameters
of the physical system. An almost complete elimination of the so-called
critical slowing down is rare, but it is quite common that
domain-decomposition methods
achieve an improved scaling behaviour, particularly so when
combined with multigrid ideas or other sophisticated techniques.

\vspace{0.2cm}\noindent
All these points are also relevant here,
although there are important differences with respect to the
well-known cases where domain-decomposition methods are used. 
In particular, the stochastic nature of 
the gauge field variables in lattice QCD 
excludes a straightforward application of multigrid methods.

%% file: sect4
\section{Algorithmic strategies for QCD}

The domain decompositions that will be considered 
in the following are divisions of the lattice 
into non-overlapping rectangular blocks such as the one shown in 
Fig.~\ref{fig1}. 
For reasons of efficiency,
the block sizes should be as large as possible,
but not larger than about $1$ fm.
More and more points are thus contained in the blocks
when the lattice spacing decreases,
while the number of blocks grows roughly in proportion
to the lattice volume.

\input figure1

The development of the simulation algorithm that
will be discussed here was guided by the general points
listed above and by further considerations
that are specific to QCD.
As already mentioned, the blocks in the block lattice 
are taken to be fairly small, with sizes less than
$1\,\fm$ or so. A strong infrared cutoff is then implied
on the fluctuations of the block fields
if Dirichlet boundary conditions are imposed
(as we shall do). In particular, since QCD is asymptotically free,
the theory becomes weakly coupled,
and therefore easy to simulate, when restricted to the blocks.
Treating the block fields separately
is thus likely to be a good strategy.

Another property of the theory that may conceivably be exploited 
is related to the fact that the effective action 
\begin{equation}
  S_{\rm eff}=\SG-2\ln\det D
  \label{eq4_1}
\end{equation}
is an approximately local functional of the SU(3) gauge potential $A_{\mu}^a$
(for simplicity a continuum notation is used in this section, where
$\SG$ denotes the Yang--Mills action and $D$ the massive Dirac operator,
and it is assumed that there are two flavours of mass-degenerate quarks).
To understand what approximate locality means and where it comes from, 
first note that, at non-zero distances $|x-y|$,
the second variation of the effective action is given by
\begin{equation}
  {\delta^2S_{\rm eff}\over\delta A^a_{\mu}(x)\delta A^b_{\nu}(y)}
  =
  2\,\tr\{T^a\dirac{\mu}S(x,y)T^b\dirac{\nu}S(y,x)\},
  \label{eq4_2}
\end{equation}
where $S(x,y)$ stands for the quark propagator in the presence of 
the gauge field.
The ellipticity of the Dirac operator implies that the quark propagator
falls off roughly like $|x-y|^{-3}$ in the short-distance regime,
and it usually decays even more rapidly at larger distances.
Practitioners in lattice QCD 
have actually long been aware of the fact that
quark propagators behave like this, for all representative gauge fields.

While this argumentation is not rigorous,
it suggests that the gauge fields in distant blocks are only weakly coupled
to each other. 
Starting with an algorithm that completely decouples the 
blocks, and treating the interactions between different blocks 
as a correction, may thus be an ansatz worth pursuing.

%% file: figure1
\begin{figure}
\centering
\epsfig{file=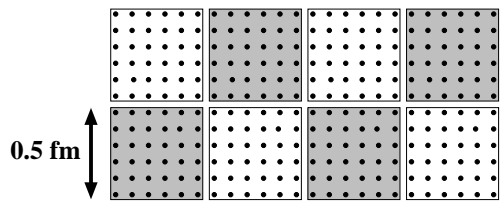,width=6.5cm,clip=true}
\caption{
Two-dimensional cross-section of 
a $24\times12^3$ lattice covered by non-overlapping $6^4$ blocks.
Only such block divisions are considered that 
can be chessboard coloured 
and where the blocks have even sizes in all dimensions.
}
\label{fig1}
\end{figure}

%% file: sect5
\section{Block decomposition of the Dirac operator}

In this and the following sections, only the standard Wilson theory will 
be considered, but all formulae generalize straightforwardly
to the O($a$) improved theory.
It is now also convenient
to set the lattice spacing to unity.
The lattice Dirac operator is 
then given by
\begin{equation}
  D=\frac{1}{2}\left\{\dirac{\mu}\left(\nabstar{\mu}+\nab{\mu}\right)
  -\nabstar{\mu}\nab{\mu}\right\}+m_0,
  \label{eq5_1}
\end{equation}
where, as usual,
$\nab{\mu}$ and $\nabstar{\mu}$ are the gauge-covariant
forward and backward difference operators and $m_0$ is 
the bare quark mass.

On a block lattice such as the one shown in Fig.~\ref{fig1},
the lattice is divided into two regions,
the union $\Om$ of all black blocks and the union $\Oms$ of all white blocks.
The exterior boundaries of these domains, 
$\dOm$ and $\dOms$, are the sets of all points at distance
$1$ from $\Om$ and $\Oms$ respectively. 

Now if the quark fields are written as a sum of two fields,
$\psi=\psi_{\Om}+\psi_{\Oms}$,
with support in $\Om$ and $\Oms$, the Dirac operator
decomposes into four pieces according to
\begin{eqnarray}
  (D\psi)_{\Om}&=&\DOm\psi_{\Om}+\DdOm\psi_{\Oms},
  \label{eq5_2}
  \\[1.5ex]
  (D\psi)_{\Oms}&=&\DdOms\psi_{\Om}+\DOms\psi_{\Oms}.
  \label{eq5_3}
\end{eqnarray}
The operator $\DOm$, for example, coincides with the Dirac operator
on the domain $\Om$, with Dirichlet boundary conditions at
$\dOm$, while $\DdOm$ is the sum of all hopping terms 
that go from $\dOm$ to $\Om$. A further decomposition
into block operators $D_{\B}$ is possible,
\begin{equation}
  \DOm=\sum_{{\rm black\;blocks}\;\B}D_{\B},
  \qquad\DOms=\sum_{{\rm white\;blocks}\;\B}D_{\B},
  \label{eq5_4}
\end{equation}
since $\Om$ and $\Oms$ are unions of disjoint blocks $\B$ that
touch one another at the edges only.

Similarly to the familiar case of even--odd preconditioning,
these equations go along with an exact factorization of the quark 
determinant,
\begin{equation}
  \det D=\prod_{{\rm all\;blocks}\;\B}\det D_{\B}
  \times\det R,
  \label{eq5_5}
\end{equation}
where $R$ is an operator that acts on the subspace
of quark fields supported on the boundary $\dOms$. If the associated
orthogonal projector is denoted by $P_{\dOms}$, the operator
is explicitly given by
\begin{equation}
  R=1-P_{\dOms}\DOm^{-1}\DdOm\DOms^{-1}\DdOms.
  \label{eq5_6}
\end{equation}
This expression looks a bit complicated, and one may have 
some doubts that the factorization (\ref{eq5_5}) 
is useful.
However, in view of the operator identity
\begin{equation}
  R^{-1}=1-P_{\dOms}D^{-1}\DdOms,
  \label{eq5_7}
\end{equation}
the solution of the equation $R\psi=\eta$, for example,
is actually no more difficult to obtain
than the solution of the Wilson--Dirac equation.

%% file: sect6
\section{Preconditioned HMC algorithm} 

Starting from the factorization (\ref{eq5_5}),
the HMC algorithm \cite{HMC}
may now be set up as usual,
with some modifications that will be explained below.
Technically speaking, the block-diagonal operator $\DOm+\DOms$
acts as a preconditioner for the HMC algorithm, with $R$ being
the preconditioned Wilson--Dirac operator.
This particular preconditioning
is actually closely related to the classical 
Schwarz alternating procedure \cite{SchwarzSolver}.

\subsection{Block decoupling}

For two mass-degenerate flavours of quarks, the pseudo-fermion action
associated to the determinants in eq.~(\ref{eq5_5}) is given by
\begin{equation}
  \Spf=\sum_{{\rm all\;blocks}\;\B}
  \left\|\Dhat_{\B}^{-1}\phi_{\B}\right\|^2+
  \left\|R^{-1}\chi\right\|^2.  
  \label{eq6_1}
\end{equation}
In this expression the block operators $D_{\B}$ have been replaced by
their even--odd preconditioned form $\Dhat_{\B}$ (which is permissible
since the determinants are the same).
The fields $\phi_{\B}$ thus reside on the even sites in the blocks $\B$
and $\chi$ is supported on $\dOms$.

As explained before, one of the algorithmic strategies
is to decouple the gauge-field variables in different blocks
as much as possible. A partial decoupling is 
easily achieved by updating only a subset 
of all gauge-field variables in each update cycle.
With respect to the field variables residing on the links
shown in Fig.~\ref{fig2}, for example, the gauge action
decomposes into a sum of 
block actions, and the same is trivially true for the block terms
in eq.~(\ref{eq6_1}). 
The field variables in different blocks
are then still coupled to each other through the 
last term in the pseudo-fermion action (the one involving
the operator $R$), but as will become clear shortly, 
this interaction is relatively weak.

\input figure2

Evidently, an algorithm that updates always the same subset of
field variables is not ergodic, but this problem
can be overcome by alternating between different block
divisions or simply by applying a random 
space--time translation to the gauge field 
after every update cycle. In the course of the simulation
all link variables are then treated equally.

\subsection{Molecular-dynamics forces}

Following the standard rules, the 
trajectories in field space 
are obtained by solving the mole\-cu\-lar-dynamics equations
\begin{eqnarray}
  {\rmd\over\rmd t}\,U(x,\mu)&=&
  \Pi(x,\mu)U(x,\mu),
  \label{eq6_2}
  \\[1.5ex]
  {\rmd\over\rmd t}\,\Pi(x,\mu)&=&
  -\Fgauge(x,\mu)-\Fblock(x,\mu)-\Fglobal(x,\mu),
  \label{eq6_3}
\end{eqnarray}
for the gauge-field variables $U(x,\mu)$ on the active links
and their canonical momenta $\Pi(x,\mu)$.
The forces $\Fgauge$, $\Fblock$ and $\Fglobal$ in these equations
derive from the gauge action, 
the block terms and the last
term in the pseudo-fermion action (\ref{eq6_1}), respectively.

\input figure3

An important observation is now that these forces have
widely different magnitudes. This is so
on all lattices considered so far, independently of the quark masses
and of whether O($a$) corrections are included or not. 
Typically the situation is as shown in Fig.~\ref{fig3}, where $\Fgauge$ is about
$4$ times larger than $\Fblock$, which is more than $6$ times larger
than $\Fglobal$. Moreover, even though the bare current-quark mass 
ranged from $14$ to $70$ MeV in this study, there is hardly any
mass dependence seen on the scale of the plot.

To some extent, at least,
these empirical findings can be understood theoretically
(cf.~Sect.~4).
The block forces $\Fblock$, for example, are protected from
large fluctuations and a significant quark-mass dependence, 
because there is a strong infrared cutoff
on the spectrum of the block Dirac operators.
Somewhat less obvious is perhaps the suppression of
the third force
\begin{equation}
  \Fglobal(x,\mu)=
  2\,\Re\!\left(R^{-1}\chi,D^{-1}
  \delta^U_{x,\mu}D
  D^{-1}D_{\dOms}\chi\right).
  \label{eq6_4}
\end{equation}
This formula shows, however, that the force
involves two quark
propagators that go from the boundaries of the blocks 
to the point $x$ where the force is to be evaluated.
The decay properties
of the quark propagator then suggest that the force should be
small, particularly so at the points $x$ that are well inside the blocks.

\subsection{Integration scheme}

The widely used numerical integration schemes for the molecular-dynamics
equations divide the integration range $0\leq t\leq\tau$ into $N$ 
steps of size $\epsilon=\tau/N$, and update the momenta
and the link variables
step by step.
Evidently the computational cost of the integration 
(and thus of the simulation) is inversely proportional to the step size.
 
When the evolution is driven by several forces of different magnitudes, 
as in the present case, it is possible to use larger step sizes 
for the smaller forces. This was first pointed out by
Sexton and Weingarten \cite{SextonWeingarten} many years ago,
and the idea was more recently revived by
Peardon and Sexton \cite{PeardonSexton}, who suggested 
to split the forces into high- and
low-frequency parts, where the latter are integrated at a lower
speed. 

Here the multiple step-size
integration method can be applied straightforwardly
since the total force is already split in parts.
It seems reasonable to choose the associated step sizes
$\epsgauge,\epsblock,\epsglobal$ such 
that 
\begin{equation}
  \epsgauge\|\Fgauge\|\simeq
  \epsblock\|\Fblock\|\simeq
  \epsglobal\|\Fglobal\|.
  \label{eq6_5}
\end{equation}
This rule works well in practice, and in all simulations
reported later the step sizes were as in Fig.~\ref{fig4}.
In this way the computational cost per trajectory is 
potentially reduced by 
a large factor, because the block-interaction force $\Fglobal$, 
which  is by far the most difficult 
to calculate at small quark masses,
is now also the least often computed force.

\input figure4

%% file: figure2
\begin{figure}
\centering
\epsfig{file=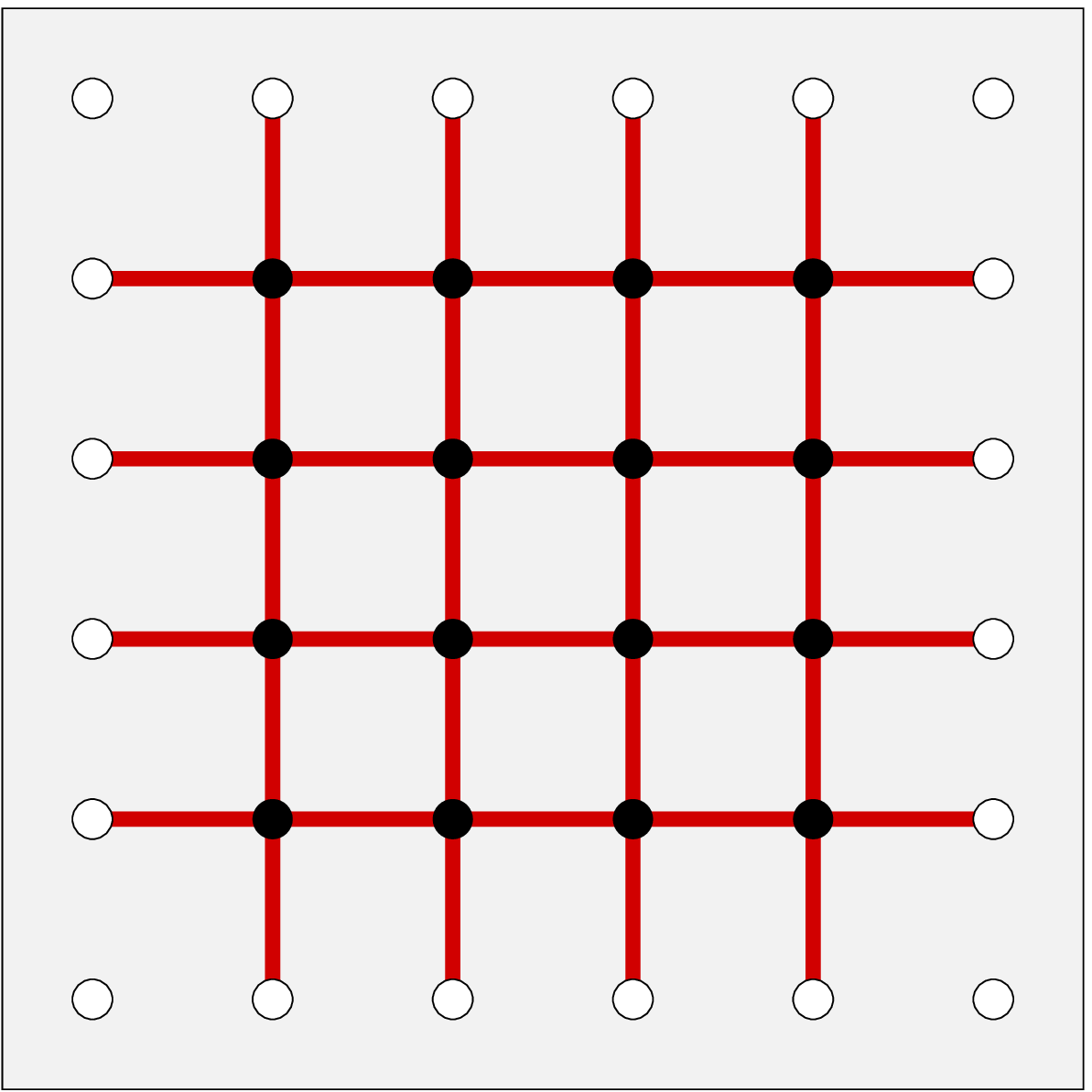,width=3.0cm,clip=true}
\caption{%
Along the molecular-dynamics trajectories, only the field variables
on the active links (thick lines) in the blocks are evolved, while
all other link variables are kept fixed. 
In this way a partial decoupling of the gauge-field variables 
in different blocks is achieved.
}
\label{fig2}
\end{figure}

%% file: figure3
\begin{figure}
\centering
\epsfig{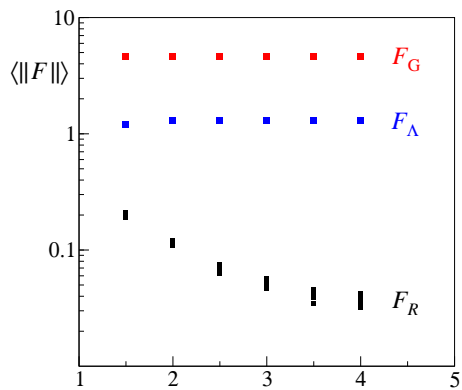}
\caption{%
Average magnitude of the forces $F_G$, $F_{\B}$ and 
$F_R$ at the end of the molecular-dynamics trajectories
as a function of the distance from the block boundary
(simulation of the O($a$) improved theory
on a $32\times16^3$ lattice divided into $8^4$ blocks).
Only $F_R$ shows a non-negligible dependence on the quark mass
(lowest four points at each distance).
}
\label{fig3}
\end{figure}

%% file: figure4
\begin{figure}
\centering
\epsfig{file=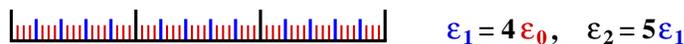,width=9.0cm,clip=true}
\caption{%
Division of the integration interval into steps of size $\epsgauge$
(small tick marks). The multiple step-size integration requires
the gauge force $\Fgauge$ to be computed at all ticks, 
the block forces $\Fblock$ at the middle-size ticks and 
the block-interaction force $\Fglobal$ at the large ticks only.
}
\label{fig4}
\end{figure}

%% file: sect7
\section{Simulations of two-flavour QCD}

Numerical studies of
the Schwarz-preconditioned HMC algorithm started about one and a half
years ago, using $8$ nodes of a PC cluster
at the Institute for Theoretical Physics at Berne \cite{SchwarzAlgorithmII}.
The algorithm is now well understood and 
more extensive simulations, on a cluster with $64$ nodes at the Fermi Institute
in Rome,
are being performed by a collaboration based at CERN and 
the University of Rome ``Tor Vergata'' \cite{CHQCD}.

The parameters of the simulations carried out so far  
are listed in Table~\ref{tab1}.
In all cases the standard Wilson theory
was considered and
the trajectory length $\tau$ was set to $0.5$.
Along the trajectories
the lattice Dirac equation on the blocks
and the full lattice was solved accurately enough 
to guarantee the reversibility of the molecular-dynamics
evolution at a level better than $10^{-9}$.
For the solution of the Dirac equation on the full lattice, a highly parallel
Schwarz-preconditioned solver was used 
\cite{SchwarzAlgorithmI}.

\input table1

In physical units the lattice spacing 
is estimated to be about $0.080$ fm at $\beta=5.6$ 
and $0.064$ fm at $\beta=5.8$ \footnote[1]{%
To some extent the quoted values for the lattice spacing 
depend on the conventions
adopted for the conversion to physical units.
Here the lattice spacing was determined, at
the specified values of the gauge coupling, 
by setting the Sommer scale $r_0$ \cite{SommerScale} to $0.5$ fm at
the quark mass where $r_0\mpi=1.26$.}.
All simulated lattices thus have
a spatial size close to $2$ fm, and a comparison
of the simulation results obtained on the $32\times24^3$ 
and the $64\times32^3$ lattice 
should therefore provide some clues as to
how large the lattice effects are at these lattice 
spacings.
Further simulations on the big lattice are still required, 
however, to match the range of quark masses covered on the small one.

An important experience made in these simulations is that 
the number $\tau/\epsglobal$
of the integration steps at which the block-interaction force $\Fglobal$ 
must be computed can be taken to be fairly small.
Moreover, to keep the acceptance rates high when the quark mass
is lowered, a moderate increase in the step numbers turns out to be sufficient.
The performance
of the algorithm will be discussed later, but
these low numbers and the fact that the simulations
could be performed on relatively small computers already show that
the algorithm works very well.

%% file: table1
\begin{table}[b]
\small
\centering
\renewcommand{\arraystretch}{1.3}
\tabcolsep0.4cm
\begin{tabular}{cccccccc}
Lattice & $\beta$ & $\kappa$ & $\sim\mq/\ms$ &
Block size & $\tau/\epsglobal$ & $N_{\rm traj}$ &
$P_{\rm acc}$\\[0.0ex]
\hline\hline
\noalign{\vskip0.5ex}
   $32\times24^3$ &
   $5.6$ &
   $0.15750$ &
   $0.93$ &
   $8\times6^2\times12$ &
   $5$ &
   $6400$ &
   $0.80$ \\
   &
   &
   $0.15800$ &
   $0.48$ &
   &
   $6$ &
   $9500$ &
   $0.80$ \\
   &
   &
   $0.15825$ &
   $0.30$ &
   &
   $10$ &
   $9400$ &
   $0.86$ \\
   &
   &
   $0.15835$ &
   $0.17$ &
   &
   $16$ &
   $5000$ &
   $0.91$ \\
   $64\times32^3$ &
   $5.8$ &
   $0.15410$ &
   $0.75$ &
   $16\times8^3$ &
   $8$ &
   $5000$ &
   $0.86$ \\
   &
   &
   $0.15440$ &
   $0.38$ &
   &
   $10$ &
   $5050$&
   $0.89$ \\
\noalign{\vskip0.5ex}
\hline\hline
\end{tabular}
\caption{%
Simulations were performed at several values of the inverse gauge
coupling $\beta=6/g_0^2$
and the hopping parameter $\kappa=(8+2m_0)^{-1}$. 
The lowest current quark mass $m$ reached in these runs
was about a sixth of the physical strange-quark mass $\ms$. 
In the last two columns, the number of trajectories generated
after thermalization and the average acceptance rate are reported.
}
\label{tab1}
\end{table}

%% file: sect8
\section{Pion mass and decay constant}

Some of the simulations reported here were performed 
at fairly low quark masses,
and the question may now be asked whether contact
with chiral perturbation theory can be made at these points.
In the case of the pion mass $\mpi$, for example, 
two-flavour chiral perturbation theory
predicts
\begin{eqnarray}
   \mpi^2&=&M^2R_{\pi},\qquad M^2\equiv2Bm,
   \label{eq8_1}  
   \\[1.5ex] 
   R_{\pi}&=&1+{M^2\over32\pi^2 F^2}\ln(M^2/\Lambda_{\pi}^2)+\ldots,
   \label{eq8_2}
\end{eqnarray}
where $m$ denotes the current quark mass, $F$ the pion decay constant
in the chiral limit, $B$ another constant related to the chiral
condensate, and $\Lambda_{\pi}$ a scale that characterizes the
strength of the one-loop chiral corrections in this formula 
\cite{GasserLeutwyler}.

\input figure5

In lattice QCD the pion mass and the current quark mass 
are usually extracted 
from the correlation functions of the isovector axial current 
and density. The data points obtained in this way on the $32\times24^3$
lattices are plotted in Fig.~\ref{fig5}. 
They all lie on a straight line, which passes through 
the origin within errors, and so it seems that the data 
leave no room for chiral loop corrections.

\input figure6

This is not obviously so, however, because 
the loop corrections could be very small.
In real-world QCD, for example,
the scale $\Lambda_{\pi}$ is expected to be such that
\cite{GasserLeutwyler} 
\begin{equation}
  \left.\ln(\Lambda_{\pi}^2/M^2)\right|_{M=140\,{\rm MeV}}
  \simeq2.9\pm2.4.
  \label{eq8_3}
\end{equation}
Using this estimate and setting $F=88$ MeV,
the one-loop expression (\ref{eq8_2}) for the correction factor $R_{\pi}$
turns out to be practically constant in the relevant range of pion masses
(see Fig.~\ref{fig6}). The pion mass squared is thus predicted to be
a nearly linear function of the quark mass in this range,
although with a slope about $4\%$ lower than $2B$.
It is then not too surprising that one-loop chiral perturbation theory
fits the data plotted in Fig.~\ref{fig5} 
if the point at the largest mass is omitted. 
Moreover, the values of the fit parameters  
$B$ and $\Lambda_{\pi}$ come out to be close to the phenomenologically
expected ones.

The situation in the case of the pion decay constant $\Fpi$ 
is qualitatively similar, 
but the one-loop correction in the chiral expansion
\begin{equation}
  \Fpi=F\left\{1-{M^2\over16\pi^2 F^2}\ln(M^2/\Lambda_F^2)+\ldots\right\}
  \label{eq8_4}
\end{equation}
tends to be larger than in eq.~(\ref{eq8_2}) by a factor $4$ or so. 
On the lattice the data for the decay constant must be renormalized before
they can be compared with this formula. 
Since the renormalization factor $Z_A$ is currently not known,
it was estimated using ``tadpole-improved'' perturbation theory.
Moreover, a small finite-volume correction, computed to
one-loop order of chiral perturbation theory \cite{GasserLeutwylerFiniteVolume},
was applied to the data.

\input figure7

Within statistical errors, 
the corrected data shown in Fig.~\ref{fig7} 
lie on a straight line. The data are also compatible with  
the chiral formula (\ref{eq8_4}) 
if the point at the largest mass is discarded and 
if $F$ and $\Lambda_F$ are treated as free fit parameters.
As it turns out, the fit result for the latter comes out to be 
well within the range 
\begin{equation}
  \left.\ln(\Lambda_F^2/M^2)\right|_{M=140\,{\rm MeV}}
              \simeq4.6\pm0.9
  \label{eq8_5}
\end{equation}
quoted by Gasser and Leutwyler \cite{GasserLeutwyler}, while 
the value obtained for the decay constant after extra\-polation to 
the physical point,
$\left.\Fpi\right|_{M=140\,{\rm MeV}}=80(7)$ MeV,
is a bit lower than expected in real-world QCD. 
This result should not be taken too seriously, however, since
there are still many uncontrolled systematic errors in 
this calculation. In particular, there can be large lattice-spacing
effects.

The conclusion is then that the simulation results for $\mpi$ and $\Fpi$
obtained on the $32\times24^3$ lattice are compatible with 
one-loop chiral perturbation theory, within errors and 
up to pion masses of about $500$ MeV. On the other hand, 
since the data can also be fitted by straight lines,
the presence of the chiral logarithms has not 
been unambiguously demonstrated. The discussion also showed, however,
that their effects are very small,
partly because of accidental numerical cancellations.
More extensive simulations will therefore be required if 
chiral perturbation theory is to be matched at this level of precision.

%% file: figure5
\begin{figure}
\centering
\epsfig{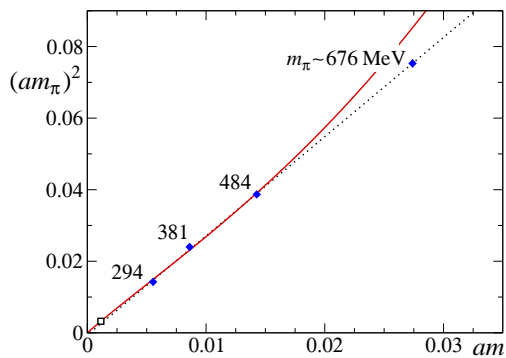}\hspace{0.45cm}
\caption{%
Simulation results for the pion mass $\mpi$ on the $32\times24^3$ lattice,
plotted versus the current quark mass $m$ in lattice units.
The dotted line is a linear fit of all four data points. 
Below  $\mpi\simeq500\,\MeV$,
one-loop chiral perturbation theory fits the data equally well
(full line). The point represented by an open square is obtained 
by extrapolation to the physical pion mass.
}
\label{fig5}
\end{figure}

%% file: figure6
\begin{figure}
\centering
\epsfig{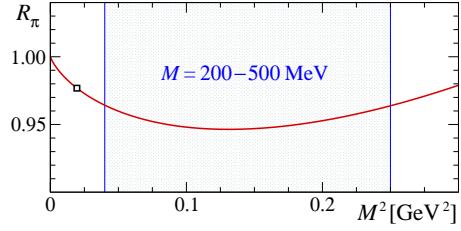}
\caption{%
Plot of the one-loop chiral correction factor (8.2)
as a function of $M^2=2Bm$, using the values 
of $F$ and $\Lambda_{\pi}$ quoted by
Gasser and Leutwyler \cite{GasserLeutwyler}.
In the shaded range, $R_{\pi}$ changes by less than $1\%$
and $\mpi^2$ is practically a linear function of the quark mass.
}
\label{fig6}
\end{figure}

%% file: figure7
\begin{figure}
\centering
\epsfig{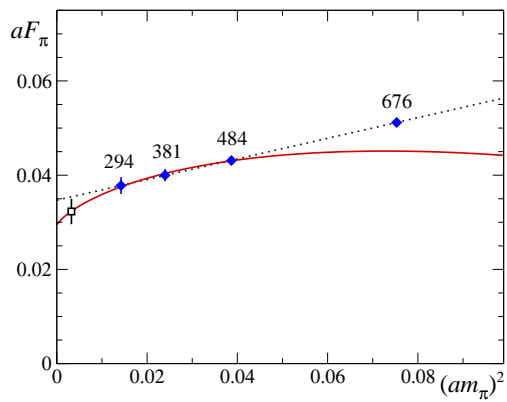}\hspace{0.0cm}
\caption{%
Simulation results for the pion decay constant $\Fpi$ on the 
$32\times24^3$ lattice,
plotted versus the pion mass squared in lattice units.
The curves are linear and chiral fits of the data as in Fig.~5.
}
\label{fig7}
\end{figure}

%% file: sect9
\section{Large-lattice experiences}

As already mentioned, the simulations  
of the $64\times32^3$ lattice performed so far
will have to be extended to smaller quark masses
to match the range of masses covered on the 
smaller lattice.
There are, however, a number of interesting 
remarks that can be made at this point.

To a first approximation, the simulation algorithm performs equally
well on the two lattices. In particular, good acceptance rates were
achieved with similarly low step numbers $\tau/\epsglobal$.
The larger volume of the big lattice actually appears to have
a stabilizing effect on the simulations
(see fig.~\ref{fig8}). 
The deficit $\Delta H$ of the molecular-dynamics Hamiltonian at the 
end of the trajectories, for example, remained bounded essentially
between $-1$ and $+1$ in all cases, and the iteration 
numbers $N_{\rm GCR}$ of the solver used for the full-lattice Dirac 
equation fluctuated with standard deviations of $3$ to $5\%$ only.

\input figure8

In general, simulating large lattices pays off in many ways.
A common experience is that 
the \emph{a priori} statistical errors tend to be smaller on these lattices
and that a sig\-ni\-fi\-cant further reduction in the errors is seen when
the correlation functions of interest are averaged over several 
distant source points.
Moreover the time dependence of the correlation functions
can be followed up to larger time separations, 
which is particularly helpful when the masses of the 
vector mesons, the baryons and other heavy particles
are to be calculated
(see Fig.~\ref{fig9}).

Note, incidentally, that the mass gap $W_{\rho}$ 
in the $\rho$-channel does not necessarily coincide
with the mass of the $\rho$-resonance. 
The virtual mixing with the isovector two-pion states actually
sets in as soon as the gap is greater than
$2\mpi$, and this leads to volume-dependent energy shifts 
that can be fairly large
\cite{Resonances}.

%% file: figure8
\begin{figure}
\centering
\epsfig{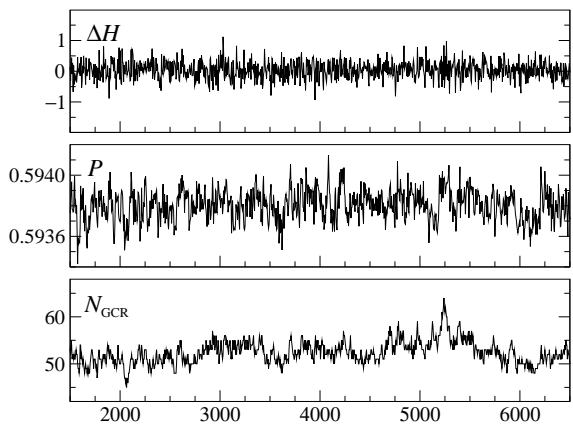}\hspace{0.0cm}
\caption{%
Histories
of the energy deficit $\Delta H$, the average plaquette $P$ and 
the average solver iteration number $N_{\rm GCR}$
as functions of the trajectory number ($64\times32^3$ lattice, 
simulation parameters as given on the last line of Table~1).
Only the values at every 5th trajectory are plotted.
}
\label{fig8}
\end{figure}

%% file: sect10
\section{Performance figures and timings}

Contrary to what may be assumed, 
the performance of a simulation algorithm is not
readily established and comparisons of different algorithms may
easily be misleading. There are a number of 
points that should be taken into account:

\begin{itemize}
\item[(a)]{In practice the efficiency of an algorithm depends on the lattice
parameters, the program and the computer. A particular algorithm may be
a good choice for simulations of coarse lattices, for example, but less 
competitive at smaller lattice spacings.}
\item[(b)]{Modern algorithms tend to have many tunable parameters
that must be properly adjusted, on each lattice, to achieve a good efficiency. 
Performing many simulations for tuning purposes alone may not be possible,
however, and good algorithms should thus have a transparent and soft 
parameter dependence.}
\item[(c)]{The relevant autocorrelation times are often
difficult to determine reliably. Moreover, when the available
data series are not exceedingly long, the calculation of autocorrelation
times invariably involves a certain amount of subjective judgement.}
\end{itemize}

\noindent
In particular, 
saying that algorithm $A$ is five times faster
than algorithm $B$, without further qualifications, 
is almost certainly a meaningless statement.

\input figure9

When a simulation algorithm spends most of the time
on the solution of the Dirac equation on the full lattice,
as is the case here, a useful
measure of the computational cost 
is the number of times the solver has to be called
before the next statistically independent field 
configuration is obtained.
A cost figure of this kind is 
\begin{equation}
  \quad\nu=10^{-3}\left(2N_2+3\right)\tau_{\rm int}[P],
  \qquad N_2=\tau/\epsglobal,
  \label{eq10_1}
\end{equation}
where $\tau_{\rm int}[P]$ denotes the integrated autocorrelation
time of the plaquette. Evidently $\nu$ is 
independent of the solver, the program and the computer
that are used for the simulation.

\input table2

Within errors, the cost figures listed in Table~\ref{tab2}
do not show any obvious dependence 
on the quark mass or the lattice size.
Moreover,
whether O($a$) improvement is switched on or not
does not appear to make a big difference.
An exceptional case are the two runs on the $64\times32^3$ lattice, 
where the cost figures reflect the high degree of stability of 
the simulations mentioned before.

Much larger values of $\nu$ are common to all
previous simulations of two-flavour QCD.
In a study of the O($a$) improved theory conducted
by the UKQCD collaboration \cite{UKQCDlight}, for
example, using the even--odd preconditioned HMC algorithm, the cost
figure on a $32\times16^3$ lattice at $\mpi\simeq420$ MeV was $5.5(11)$.
Significantly smaller masses were reached by the
CP--PACS collaboration on a coarse $24\times12^3$ lattice
\cite{CPPACSsmall}, and in these simulations, $\nu$ ranged from
$9.4(18)$ at $\mpi\simeq398$ MeV to $29(10)$ at $\mpi\simeq255$ MeV.
From this purely algorithmic point of view, the new algorithm
is thus much more efficient than the algorithms previously used,
particularly so at small quark masses.

\input figure10

In practice the computer time required for a particular simulation
also depends on the specific capabilities
of the available computers and the efficiency of the program.
So far the algorithm has been implemented on PC clusters,
and it performs very well on these machines 
(see Fig.~\ref{fig10}).
Moreover, the timings show that the algorithm scales
favourably with the quark mass and the lattice size,
a property that should be largely independent of 
the computer and the program.

%% file: figure9
\begin{figure}
\centering
\epsfig{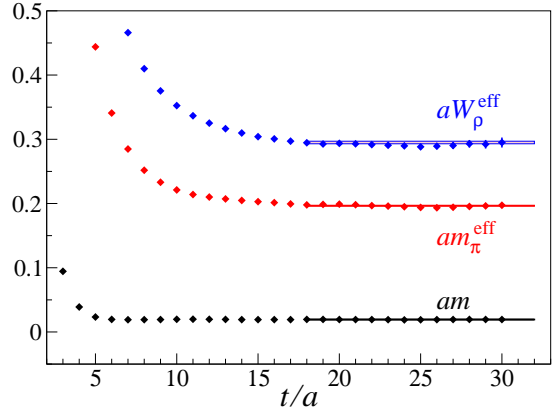}\hspace{0.0cm}
\caption{%
Plot of the current-quark mass $m$, the pion mass $\mpi$ and the 
mass gap $W_{\rho}$ in the $\rho$-meson channel
($64\times32^3$ lattice, 
simulation parameters as given on the next-to-last line of table~1).
The values shown are effective masses in lattice units, 
obtained from the appropriate
correlation functions at times $t$ and $t-a$.
}
\label{fig9}
\end{figure}

%% file: table2
\begin{table}[b]
\small
\centering
\small
\newdimen\digitwidth
\setbox0=\hbox{\rm 0}
\digitwidth=\wd0
\catcode`@=\active
\def@{\kern\digitwidth}
\renewcommand{\arraystretch}{1.15}
\tabcolsep0.5cm
\begin{tabular}{cccccc}
Lattice & $\beta$ &$c_{\rm sw}$ & $am$ & 
$\tau_{\rm int}[P]$ & $\nu$\\[0.0ex]
\hline\hline
\noalign{\vskip0.5ex}
   $32\times16^3$ &
   $5.3$ &
   $1.9095$ &
   $0.035$ &
   $56(26)$ &
   $0.84(39)$ \\
   &
   &
   &
   $0.021$ &
   $14(4)@$ &
   $0.27(8)@$ \\
   &
   &
   &
   $0.011$ &
   $21(6)@$ &
   $0.40(11)$ \\
   &
   &
   &
   $0.007$ &
   $17(5)@$ &
   $0.39(12)$ \\
\noalign{\vskip0.5ex}
   $32\times24^3$ &
   $5.6$ &
   $0.0$ &
   $0.027$ &
   $53(22)$ &
   $0.69(29)$ \\
   &
   &
   &
   $0.014$ &
   $33(11)$ &
   $0.50(17)$ \\
   &
   &
   &
   $0.009$ &
   $27(10)$ &
   $0.62(23)$ \\
   &
   &
   &
   $0.006$ &
   $21(5)@$ &
   $0.74(18)$ \\
\noalign{\vskip0.5ex}
   $64\times32^3$ &
   $5.8$ &
   $0.0$ &
   $0.019$ &
   $16(3)@$ &
   $0.30(6)@$ \\
   &
   &
   &
   $0.011$ &
   $13(2)@$ &
   $0.30(5)@$ \\
\noalign{\vskip0.5ex}
\hline\hline
\end{tabular}
\caption{%
Estimates of the integrated autocorrelation time of the 
plaquette (in numbers of trajectories of length $\tau=0.5$)
and of the cost figure $\nu$ [eq.~(10.1)]. 
The results listed on the first few lines refer to the 
O($a$) improved theory, with non-perturbatively tuned
coeffcient $c_{\rm sw}$ of the Sheikholeslami--Wohlert term \cite{csw}. 
}
\label{tab2}
\end{table}

%% file: figure10
\begin{figure}
\centering
\epsfig{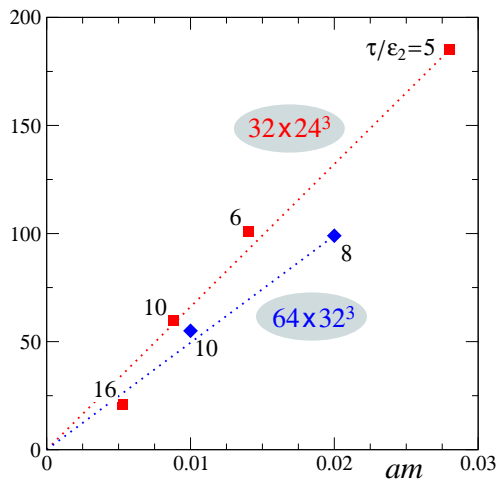}\hspace{0.0cm}
\caption{%
Number of accepted gauge-field configurations generated per day
at the simulation points listed in table~1. 
All simulations were performed on 
commodity PC clusters, using $8$ and $32$ nodes, respectively,
for the $32\times24^3$ and $64\times32^3$ lattices.
}
\label{fig10}
\end{figure}

%% file: sect11
\section{Conclusions}

The important qualitative message of this talk is 
that numerical simulations of lattice QCD with light Wilson quarks
are much less ``expensive'' than previously estimated. 
Lattices as large as $96\times48^3$, for example,
with lattice spacings $a=0.06$--$0.08$ fm and at pion masses
from, say, $200$ to $500$ MeV,
should now be accessible
using a (current) PC cluster with a few hundred nodes.
As a result contact with chiral perturbation theory 
can be made at lattice spacings where the residual lattice
effects are expected to be small.

After having passed many tests successfully,
the use of the Schwarz-preconditioned HMC al\-go\-rithm 
can be recommended without reservation if 
large lattices, with lattice spacings $a\leq0.1\,\fm$,
are to be simulated. First studies now also confirm
that the efficiency of the algorithm is preserved when
O($a$) improvement is switched on. There appears to be
very little difference here, and it also seems unlikely 
that the addition of six-link terms 
to the gauge action \cite{IwasakiI}--\cite{LWactionII}
will slow down the simulations significantly. 
Some thought may have to be given, however, 
to which is the best way to include the strange sea quark,
although this quark is so heavy that one may be able to do
with a preconditioned PHMC algorithm \cite{PHMCI,PHMCII}.

Not many physics issues were touched in this talk,
but there is a wide range of topics that may now be addressed
in a conceptually solid framework, including 
pion scattering, the $\rho$-resonance, the properties of the nucleons
and the charmed mesons, to mention just the obvious ones.
Sometimes it seems that the most
difficult question in these days is: What should we compute first? 

\vspace{0.3cm}
Many results presented here were obtained together
with Luigi Del Debbio, Leo\-nar\-do Giusti, Roberto Petronzio and
Nazario Tantalo, whom I would like to thank for 
helpful discussions and a pleasant collaboration.
The numerical simulations were performed on a PC cluster at
the Institut f\"ur Theo\-retische Physik der Universit\"at Bern,
which was funded in part by
the Schweizerischer Nationalfonds, and on another PC cluster
at the Fermi Institute in Rome.
I wish to thank all these institutions for their support.